# Interference Alignment: A one-sided approach

Hadi G. Ghauch
*Information Networking Institute*
Carnegie Mellon University
Pittsburgh, PA, USA
hghauch@cmu.edu

Constantinos B. Papadias
*Broadband Wireless & Network Sensors Lab*
Athens Information Technology
Athens, Greece
papadias@ait.gr

*Abstract*—Interference Alignment (IA) is the process of designing signals in such a way that they cast overlapping shadows at their unintended receivers, while remaining distinguishable at the intended ones [1]. Our goal in this paper is to come up with an algorithm for IA that runs at the transmitters only (and is transparent to the receivers), that doesn't require channel reciprocity, and that alleviates the need to alternate between the forward and reverse network as is the case in [2], thereby inducing significant overhead in certain environments where the channel changes frequently. Most importantly, our effort is focused on ensuring that this one-sided approach does not degrade the performance of the system w.r.t. [2] (since it cannot improve it). As a first step, we model the interference in each receiver's desired signal as a function of the transmitters' beamforming vectors. We then propose a simple steepest descent (SD) algorithm and use it to minimize the interference in each receiver's desired signal space. We mathematically establish equivalences between our approach and the Distributed IA algorithm presented in [2] and show that our algorithm also converges to an alignment solution (when the solution is feasible).

## I. INTRODUCTION

Interference Alignment (IA) is a recent technique that proved the achievability of the sum-rate capacity of the *K*-user MIMO Interference Channel (IC),

$$C_{SR} = (KM/2)\log(1+P) + o(\log(P)) \qquad (1)$$

(first presented implicitly in [4] and then further elaborated in [1], [5]). The key to the achievability of this result is forcing each transmitter to use only half of its signaling space, and each receiver to partition its received space into two equally sized subspaces: one is intended for the desired signal, while the other is left for the interference [3]. As a result, each transmitter-receiver pair is able to communicate over an interference-free space, irrespective of the number of interferers.

In a nutshell, the problem of IA boils down to finding transmit precoding ($V^{[1]},...,V^{[K]}$) and receive interference suppression matrices ($U^{[1]},...,U^{[K]}$), to cancel all the unwanted interference at each receiver. An elegant iterative algorithm that exploits channel reciprocity to alternate between the forward and reverse network, and find such matrices in a distributed way, was presented in [2]. However, such an approach exhibits several aspects that might be perceived as drawbacks if one wishes to apply this algorithm to certain environments. For instance, alternating between the forward and reverse network requires tight synchronization at both ends (which may be hard to achieve, especially at the receivers' side). Moreover, this alternation might induce significant overhead in a dynamic environment where the channel varies rapidly. Furthermore, the receivers, having generally limited computational complexity, might be a bottleneck for the execution time of the algorithm. And finally, the assumption of channel reciprocity practically limits the applicability of such an algorithm to TDD systems.

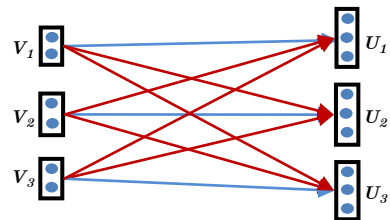

**Figure 1:** *3*-user *2x3* MIMO IC

As is the case in [2], most of the IA algorithms presented in the literature such as [6] and [7], perform the optimization of some cost function over both the precoding matrices and interference suppression filters i.e. $V^{[1]},...,V^{[K]}, U^{[1]},...,U^{[K]}$. Although such an algorithm is desirable in terms of performance, an inevitable consequence is the fact that both the transmitters and receivers are active in the algorithm, an implication that we may wish to avoid for the reasons that we previously stated. Our aim in this paper is to present an approach that decouples the IA problem, by restricting the optimization to the transmitters' side only. A clear advantage of such an approach is the fact that, as seen by the receivers, the algorithm is *transparent* (g1), i.e. all a receiver has to do is to pick the subspace with the lowest interference (not more than what is already in use). Moreover, following such an approach *bypasses the overhead* and other complications generated by alternating between the forward and reverse network (g2). Furthermore, by adopting this approach, we *relax the assumption of channel reciprocity* and make our algorithm applicable to both TDD and FDD systems (g3). We keep in mind that we do not expect our approach to outperform the results of [2], and especially other non-subspace methods [7], at low SNR. In fact, by reducing the search space from $V^{[1]},...,V^{[K]}$ to $V^{[1]},...,V^{[K]}, U^{[1]},...,U^{[K]}$ only, it is unlikely that we end up with a better solution. However, from the start, our aim was to design a simple algorithm with the above guidelines in mind (g1-g3). This said, the bulk of the effort is to ensure that this approach is feasible and does not incur a significant loss in performance.

This work has been supported by the EU FET Project HIATUS. The project HIATUS acknowledges the financial support of the Future and Emerging Technologies (FET) programme, within the Seventh Framework Programme for Research of the European Commission, under FET-Open grant number: 265578.

The rest of the paper is organized as follows. In Section II we present our signal model, in Section III we put forth the algorithm derivation, consisting of the mathematical model and of the proposed algorithm, and in Section IV we present numerical results and discuss them.

In the following, bold uppercase letters denote matrices/vectors. We assume that the eigenvalues of a matrix $A$, and their corresponding eigenvectors, are sorted in increasing order. Therefore, $\lambda_i[A]$ denotes the $i^{th}$ eigenvalue of $A$. The ' symbol denotes the transpose of a matrix, * its complex conjugate, $^H$ its conjugate transpose (Hermitian), while the operator $\langle Z, Z \rangle$ denotes the inner product of a vector/matrix. Moreover, we denote by $\mathcal{K} = \{1,...,K\}$ the set of integers from 1 to $K$. Finally, $I_n$ denotes the $n \times n$ identity matrix.

## II. SIGNAL MODEL

We build upon the model introduced in [2]. We consider a $K$-user MIMO Interference Channel (IC) where the $k^{th}$ transmitter-receiver pair is equipped with $M^{[k]}$ and $N^{[k]}$ antennas, respectively. $d^{[k]}$ is the desired number of streams between the $k^{th}$ transmitter-receiver pair, where $d^{[k]} \leq \min(M^{[k]}, N^{[k]})$. Moreover, $H^{[kj]}$ denotes the $N^{[j]} \times M^{[k]}$ channel matrix from transmitter $j$ to receiver $k$, and is assumed to have i.i.d complex Gaussian random variables, drawn from a continuous distribution. Finally, $V^{[j]}$ denotes the $j^{th}$ transmitter's $M^{[k]} \times d^{[k]}$ precoding matrix ($1 \leq j \leq K$), whose orthonormal columns span the $d$-dimensional space at the transmitter. We denote the received signal vector at receiver $k$ after interference suppression by

$$Y_s^{[k]} = U^{[k]H}Y^{[k]} = U^{[k]}(\sum_{j=1}^{K} H^{[kj]}V^{[j]}X^{[j]} + Z^{[k]}), k \in \mathcal{K} \quad (2)$$

where $U^{[k]}$ is the $N^{[k]} \times d^{[k]}$ interference suppression filter at receiver $k$, $X^{[j]}$ is the $d^{[j]} \times 1$ vector of independently encoded Gaussian symbols of transmitter $j$, with covariance matrix $(P^{[j]}/d^{[j]})I_d$, and $Z^{[k]}$ is the i.i.d complex Gaussian noise at receiver $k$ with unit variance.

## III. ALGORITHM DERIVATION

### A. Motivation

As we previously stated, the IA conditions imply that we need to make the interference at each receiver align by occupying an $(N-d)$-dimensional subspace, creating an interference-free, $d$-dimensional space for the desired signal (where $d$ must equal $N/2$ to achieve the MIMO IC capacity, i.e. $KM/2$ degrees of freedom (DoFs); only feasible when $K \leq 3$). Equivalently, if we are able to create at each receiver a $d$-dimensional signal space that is free from interference, we have implicitly aligned the interference in the other remaining space. This done, all a receiver has to do is to "hide from the interference" by projecting the received signal onto the $d$-dimensional subspace that has the lowest interference, thereby suppressing all the undesired interference (Fig. 2).

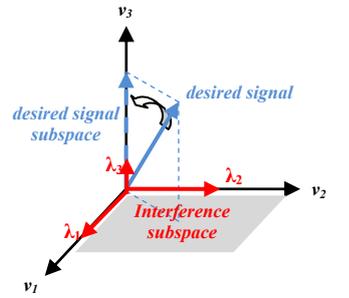

**Figure 2: 3D signal space at Rx**

In that sense, our algorithm should aim at creating a $d$-dimensional subspace that is free from interference, at each receiver. The natural question that arises is: what is a good metric (cost function) for such a purpose? We know from Principal / Minor Component Analysis that the eigenvectors of the interference covariance matrix correspond to the dimensions along which a receiver "sees" interference, and their corresponding eigenvalues indicate the variance / power of interference along that dimension (Fig. 2). Assuming that the eigenvalues and their corresponding eigenvectors are sorted in increasing order, this implies that the $d$-dimensional subspace with the lowest interference is spanned by the $d$-eigenvectors corresponding to the $d$-smallest eigenvalues, and the variance of interference in this subspace is given by the sum of the $d$-smallest eigenvalues (used in [2]). Summing up, we wish to track the variance of interference in the desired signal space, by tracking the sum of the $d$-smallest eigenvalues of the interference covariance matrix at each receiver, as the transmitters vary their precoding matrices.

### B. Problem formulation

This said, at each receiver, we seek to minimize the sum of the $d$-smallest eigenvalues of the interference covariance matrix, over the set of transmit precoding matrices $V^{[1]},...,V^{[K]}$. Thus, we define our cost function:

$$\min_{V^{[1]},...,V^{[K]}} f = \sum_{k=1}^{K} \sum_{i=1}^{d^{[k]}} \lambda_i[Q^{[k]}], k \in \mathcal{K} \quad (3)$$
$$\text{subject to } V^{[j]H}V^{[j]} = I_d, \forall, j \in \mathcal{K}$$

$Q^{[k]}$ is the $k^{th}$ receiver's interference covariance matrix:

$$Q^{[k]} = \sum_{\substack{j=1 \\ j \neq k}}^{K} \frac{P^{[j]}}{d^{[j]}} H^{[kj]}V^{[j]}V^{[j]H}H^{[kj]H} \quad (4)$$

Now that we defined the cost function, we need a way to minimize it. Intuitively, one first thinks about derivatives. However, in this case, the variables in question are matrices (we need derivatives of the form $dQ/dV$, where both $Q$ and $V$ are matrices).

### C. Mathematical Model

Although our goal is to propose an algorithm for IA, another notable feature of this work is the novel model that we derive, for modeling the interference in the network. It draws its foundations in matrix differential calculus (matrix differentials [10]). By using matrix differentials (which can be thought of as derivatives for matrices), we hope to find expressions of the form $df = DF.dV$ where $df$ and $dV$ are the differential of the function and of the variable, respectively, and where $DF$ is a Jacobian matrix (a matrix of partial derivatives) that translates infinitesimal changes in $V$ (i.e. $dV$), to changes in $f$ [10]. The reason for seeking such matrices is the fact that Jacobian matrices go hand-in-hand with numerical minimization. Moreover, differentials are easy to work with since they exhibit linearity. Due to space limitations, we omit most of the mathematical derivations and only present the results. The model consists of two building blocks that are chained together.

*1) Modeling Covariance Matrices*

We show mathematically that for a *K*-user MIMO IC, the interference covariance matrix at receiver *k* satisfies:

$$vec(dQ^{[k]}) = \sum_{\substack{j=1 \\ j \neq k}}^{K} DQ_R^{[kj]} vec(dV_R^{[j]}) + DQ_I^{[kj]} vec(dV_I^{[j]}), \forall k \in \mathcal{K} \quad (5)$$

where the $vec()$ operator stacks the columns of the operand matrix vertically; $V_R$ and $V_I$ denote the real and imaginary parts of $V$, respectively. The above equation answers the following question: *How do the entries in $Q^{[k]}$ vary ($1 \leq k \leq K$), when any element in $V^{[j]}$ changes ($1 \leq j \leq K$)?* In other words, the above equation relates the changes in any signaling vector at any transmitter, to all the changes that they induce at the corresponding covariance matrices.

*2) Modeling Eigenvalues of Covariance Marices*

Since we are interested in minimizing sums of eigenvalues (3), the second logical step would be to derive a model that relates the differential of a sum of eigenvalues of a given covariance matrix, to the differential of the covariance matrix itself. We extend a result presented in [9], and derive the following model. For a given covariance matrix $Q^{[k]}$ (real eigenvalues, since $Q^{[k]}$ is Hermitian), the differential of the sum of its *d*-smallest eigenvalues can be written as:

$$d(\lambda_S(Q^{[k]})) = (\sum_{i=1}^{d^{[k]}} DL_i^{[k]}) vec(dQ^{[k]}) = DL_S^{[k]} vec(dQ^{[k]}), \forall k \in \mathcal{K} \quad (6)$$

where $\lambda_S(Q^{[k]}) = \sum_{i=1}^{d^{[k]}} \lambda_i(Q^{[k]})$ represents the sum of the *d*-smallest eigenvalues of a covariance matrix $Q^{[k]}$, and $DL_i^{[k]}$ is the Jacobian matrix associated with the $i^{th}$ eigenvalue of $Q^{[k]}$ (here we also omit the expression for $DL_i^{[k]}$).

*3) Chaining the two blocks*

To combine the two blocks, we substitute the expression of $vec(dQ^{[k]})$ in (5) into (6), to get:

$$d(\lambda_S(Q^{[k]})) = DL_S^{[k]} \sum_{\substack{j=1 \\ j \neq k}}^{K} DQ_R^{[kj]} vec(dV_R^{[j]}) + DQ_I^{[kj]} vec(dV_I^{[j]}), k \in \mathcal{K} \quad (7)$$

where $k \in \mathcal{K}$. In this equation, w*e address the issue of how does the sum of the eigenvalues of the interference covariance matrix at each receiver (the variance of interference in the desired signal space) change, as any element in $V^{[1]},...,V^{[K]}$ changes (the transmitters' vectors)*.

Using the above equation and exploiting the linearity of differentials, we write the differential of our cost function

$$df = \sum_{k=1}^{K} DL_S^{[k]} \{ \sum_{\substack{j=1 \\ j \neq k}}^{K} (DQ_R^{[kj]} vec(dV_R^{[j]}) + DQ_I^{[kj]} vec(dV_I^{[j]}))\} \quad (8)$$

We also notice that the above expression can be conveniently placed in matrix form (shown at bottom of page)

$$df = DL^A.DQ_R^A.vec(dV_R^A) + DL^A.DQ_I^A.vec(dV_I^A)$$
$$= D\Psi_R^A vec(dV_R^A) + D\Psi_I^A vec(dV_I^A)$$
$$= \begin{bmatrix} D\Psi_R^{[1]} & ... & D\Psi_R^{[K]} \end{bmatrix} \begin{bmatrix} vec(dV_R^{[1]}) \\ : \\ vec(dV_R^{[K]}) \end{bmatrix} + \begin{bmatrix} D\Psi_I^{[1]} & ... & D\Psi_I^{[K]} \end{bmatrix} \begin{bmatrix} vec(dV_I^{[1]}) \\ : \\ vec(dV_I^{[K]}) \end{bmatrix} \quad (9)$$

where in the last two equations, the *A* superscript denotes the augmented matrix formed by concatenating all the sub-matrices together. By examining the last expression for *df*, two statements can be made. Firstly, we have achieved our goal and expressed the differential of our cost function as $df = DF.dV$ (actually, in our case, we have two Jacobian matrices). Thus, we have derived the Jacobian matrices that we need to perform our minimization. It is worth mentioning here that a direct analytical attempt to find the derivative of *f* would be tedious due to the non-linear $\lambda_i[...]$ operator in our cost function. Secondly, we note that it is easy to show that (8) is nothing but the first order linear approximation of a scalar function of a vector $x$, using the gradient $f(x) = f(x_o) + (\nabla f)_{xo}(x - x_o)$. Thus, effectively, we have transformed a seemingly complex optimization problem involving a set of *K* matrices, into a simple gradient-like expression suitable for any gradient descent algorithm.

D. *Proposed Algorithm*

As previously said, the motivation behind our model that employs differentials and Jacobian matrices, is the fact they go hand-in-hand with numerical minimization problems. The general structure of our Steepest Descent (SD) algorithm is inspired by Algorithm 15 in [8], with the extension that every variable $vec(dV^{[j]})$ has a Jacobian matrix, a steepest decent direction and a step size associated with it. Since the algorithm is intuitive, and due to the space limitations, we have omitted most of the mathematical derivations (we refer the reader to [8]). The full algorithm is presented in Table 1. Below we comment on some of its steps.

<u>*Step 1:*</u> By comparing (9) & (11), it can be easily verified that

$$D\Psi_R^{[j]} = \sum_{\substack{k=1 \\ k \neq j}}^{K} DL_S^{[k]} DQ_R^{[kj]} \text{ and } D\Psi_I^{[j]} = \sum_{\substack{k=1 \\ k \neq j}}^{K} DL_S^{[k]} DQ_I^{[kj]} \quad (10)$$

<u>*Step 3*</u> is required for steps 4 & 5. Note that the *GS{.}* operator denotes the Gram-Schmidt Orthogonalization of a matrix, and is described in detail in [11].

<u>*Steps 4,5:*</u> They can be thought of as "calibration steps" for the algorithm step size (Armijo's rule), to ensure convergence (initially presented in [12], then later used by [8]).

$$df = \begin{bmatrix} DL_S^{[1]} & ... & DL_S^{[K]} \end{bmatrix} \begin{bmatrix} 0 & ... & DQ_R^{[1K]} \\ : & & : \\ DQ_R^{[K1]} & ... & 0 \end{bmatrix} \begin{bmatrix} vec(dV_R^{[1]}) \\ : \\ vec(dV_R^{[K]}) \end{bmatrix} + \begin{bmatrix} DL_S^{[1]} & ... & DL_S^{[K]} \end{bmatrix} \begin{bmatrix} 0 & ... & DQ_I^{[1K]} \\ : & & : \\ DQ_I^{[K1]} & ... & 0 \end{bmatrix} \begin{bmatrix} vec(dV_I^{[1]}) \\ : \\ vec(dV_I^{[K]}) \end{bmatrix} \quad (11)$$

```
Start with random V^[1], ..., V^[K]
for n = 1, 2, ..
    for j = 1, ..., K
        1. Compute the corresponding Jacobian matrix DΨ_R^[j] and DΨ_I^[j]
        2. Find steepest descent direction vec(Z^[j]) = -(DΨ_R^[j] + iDΨ_I^[j])'
        3. Find P_1^[j] = GS{V^[j] + 2γ^[j]Z^[j]}
                 P_2^[j] = GS{V^[j] + γ^[j]Z^[j]}
        4. while ( f(V^[1], ..., V^[K]) - f(V^[1], ..., V^[j-1], P_1^[j], V^[j+1], ..., V^[K])
                   ≥ 2γ^[j] < Z^[j], Z^[j] > )
                 then γ^[j] = 2γ^[j], P_1^[j] = GS{V^[j] + 2γ^[j]Z^[j]}
        5. while ( f(V^[1], ..., V^[K]) - f(V^[1], ..., V^[j-1], P_2^[j], V^[j+1], ..., V^[K])
                   < γ^[j] < Z^[j], Z^[j] > )
                 then γ^[j] = γ^[j]/2, P_2^[j] = GS{V^[j] + γ^[j]Z^[j]}
        6. (V^[j])^{n+1} = (V^[j])^n + γ^[j]Z^[j]
        7. (V^[j])^{n+1} = GS{(V^[j])^{n+1}}
        8. if(| < Z^[j], Z^[j] > | < desired tolerance) then exit
    end
end
```

**Table 1: Proposed SD Algorithm**

In a nutshell, step 4 ensures that the choice of $\gamma$ will significantly reduce the function, while step 5 ensures that the step is not too big to overshoot the optimal point.

*Steps 6,7:* Step 6 is simple and intuitive. However, the newly computed solution $V^{[1]}, ..., V^{[K]}$ does not satisfy the unitary constraints. This can be effectively accomplished by using step 7 that will project the solution back onto the constraint surface (the surface of a hypershpere) [11].

*Step 8:* It was shown in [8] that
$<Z, Z> = \Re\{tr(Z^H(I - 1/2 XX^H)Z)\}$ where $X$ is the optimization variable and $Z$ is the SD direction. Exploiting the fact that $tr(A^H X) = vec(A)^H vec(X)$, we show that:
$<Z^{[j]}, Z^{[j]}> = \Re\{vec(Z^{[j]})^H (I_d \otimes (1/2)V^{[j]}V^{[j]H}) vec(Z^{[j]})\}$  (12)
where $\otimes$ denotes the Kronecker product for matrices.

## IV. NUMERICAL RESULTS

Before we move to presenting our simulations, we remind the reader that we do not expect our approach to outperform other ones (Section I). Rather, we want to ensure that this one-sided approach does not entail any significant loss in performance compared to [2].

First, we simulated a *3*-user *2x2* MIMO IC where the desired DoFs per user are *1 (d = 1)*. As we can see, in this given example (Fig. 3), after the algorithm converges, the cost function had decreased by a factor of *1.12x10^6* in *22* iterations, and remains at this level, a clear indication of the algorithm's performance. We note that more sophisticated algorithms can be used for faster and more accurate convergence, such as Newton-type methods [8] or RLS-type algorithms [15]. However, we still have to check if the final solution satisfies the IA conditions. Since we have two interfering signals at each receiver, we are able to plot the angle between the spaces spanned by each interfering signal, at each receiver. Evidently, a perfect IA solution is characterized by a zero angle at each receiver. Referring to Fig. 4, the algorithm decreases the subspace between the two interfering signals at each receiver to zero (asymptotically). We also compare the performance of this algorithm to the Distributed IA algorithm in [2]. In Fig.5, the dashed line is the theoretical sum-rate capacity predicted by IA, i.e. $C_{SR} = (KM/2)\log(1+P)$. As we can see, the curves that correspond to our algorithm and Distributed IA (with square and diamond shaped points, respectively) are almost overlapping indicating the same performance. Moreover, we notice that at high SNR, they follow the linear scaling predicted by IA, thus verifying that they indeed succeed in aligning the interference.

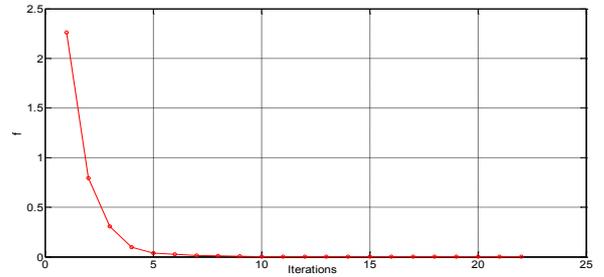

**Figure 3: Cost function**

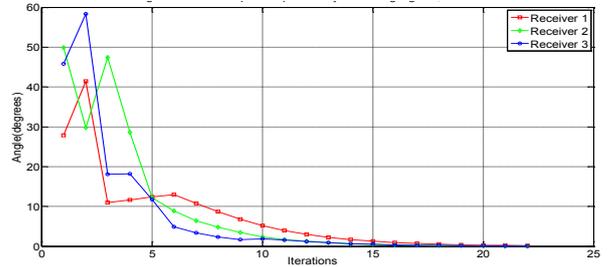

**Figure 4: Angle between spaces spanned by interfering signals, at each Rx**

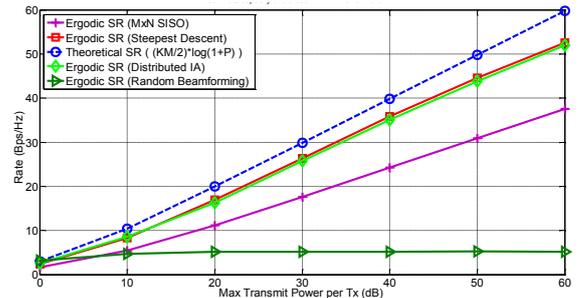

**Figure 5: Ergodic Sum-Rate (SR) capacity of a *3*-user *2x2* MIMO IC with *d = 1***

*Remark:*
*To satisfy the IA conditions, each receiver must be able to partition its received N-dimensional signal space into two subspaces of equal size (i.e., d = N/2) where each user is able to achieve M/2 DoFs (for a total of KM/2, the SR capacity of the K-user MIMO IC). However, for cases where the alignment is not feasible, i.e. K > 3, the algorithm still creates an interference-free d-dimensional subspace but with*

$d < N/2$. Thus, in this suboptimal case, each user is still able to achieve $d$ DoFs (where $0 < d < N/2$, for a total that is strictly less than $KM/2$), implying that, although at a slower rate, the capacity still scales with $log(1+SNR)$: a clear improvement over the interference limited case (i.e., $d = 0$).

## V. DISCUSSION

In this section we establish equivalences between our approach and [2]. Interestingly, it is straightforward and intuitive to mathematically verify that our cost function is nothing but the Weighted Leakage Interference (WLI) in [2] (the derivation is omitted due to the lack of space). Thus, the following points are common grounds for comparison:

- In terms of convergence, our algorithm aligns interference in a matter similar to Distributed IA, i.e. it converges to an IA solution when it is feasible (the fact that the cost function decreases with every iteration, is an inherent property of SD algorithms). However, as is the case in [2], convergence to a global optimum is not guaranteed, because even though our cost function might be convex under certain assumptions, the unitary constraints by themselves form a non-convex set.
- Simulations show that there is no loss in capacity. This result is further backed by the fact that our approach and [2] are both subspace methods, i.e. they both ignore noise when finding the subspace with the least interference (optimal at high SNR only).

However, in a realistic setting, the following are seen as clear benefits of our approach:

- We removed $\boldsymbol{U}^{[1]},...,\boldsymbol{U}^{[K]}$ from the minimization. Thus, the algorithm can run at the transmitters only, getting around the need to alternate between forward and reverse network (an inherent property of alternating minimization algorithms) which can induce a significant overhead in mobile environments.
- The algorithm is completely transparent to the receivers, i.e. they are oblivious to the entire process. All they have to do is "tune" to the subspace that has the lowest interference (not more than what is already available). The purpose of this entire approach is to ensure that this subspace is interference-free

We note that a typical drawback of such IA techniques is the need for global channel knowledge at the transmitters. However, this assumption is not unrealistic since recent results such as [13] and [14] show that under certain mild assumptions, limited feedback can achieve the full sum-rate degrees-of-freedom of the *K*-user MIMO/SISO IC.

## VI. CONCLUSIONS AND FUTURE WORK

To sum up, we derived a model to track the interference in the desired signal space of each receiver and used it to perform IA, on the transmitters' side solely. As a proof of concept of the validity of our approach, we applied a simple SD algorithm to it, and showed that it converges to an IA solution (when feasible). Moreover, in contrast to other proposed algorithms, our approach decouples the IA problem: instead of both the transmitters and receivers participating, we get around the need to alternate between transmitters and receivers by shifting the computational involvement to the transmitters' side, making it transparent to the receivers (an improvement over existing IA algorithms), without incurring any apparent loss in the algorithm performance. In the future, we will try using more advanced algorithms for faster convergence. Furthermore, we wish to investigate the possibility of finding conditions under which we ensure that the algorithm converges to a global optimum (by relaxing some constraints to ensure convexity).

### ACKNOWLEDGMENT

The authors would like to acknowledge the many helpful discussions with Professor Spyridon Vassilaras, of AIT's B-WISE Lab.

### REFERENCES

[1] V. R. Cadambe and S. A. Jafar, "Interference alignment and degrees of dreedom of the K-user interference channel" *IEEE Trans. Information Theory*, vol. 54, no. 8, pp. 3425–3441, Aug. 2008

[2] K. Gomadam, V. R. Cadambe, and S. A. Jafar, "Approaching the capacity of wireless networks through distributed interference alignment", available on arXiv.

[3] Cadambe, Jafar, "Reflections on interference alignment and the degrees of freedom of the K-user MIMO interference channel", *IEEE Inform Theory Society Newsletter*, Vol. 54, No. 4, Dec. 09, pp. 5 - 8.

[4] M. Maddah-Ali, A. Motahari, and A. Khandani, "Signaling over MIMO multi-base systems—Combination of multi-access and broadcast schemes" in *Proc. Int. Symp. Inform Theory 2006*, pp. 2104–2108.

[5] M. A. Maddah-Ali, A. S. Motahari, and A. K. Khandani, "Communication over MIMO X channels: Interference alignment, decomposition, and performance analysis" *IEEE Trans. Inform. Theory*, vol. 54, no. 8, pp. 3457–3470, Aug. 2008.

[6] S. W. Peters, R. W. Heath, "Interference alignemnt via altenating minimization", *Proc. of IEEE Int. Conf. on Acoustics, Speech, and Signal Processing*, Taipei, Taiwan, April 2009, pp. 2445 – 2448.

[7] S. W. Peters, R. W. Heath, Jr, "Cooperative algorithms for MIMO interference channels" , to appear in *IEEE Trans. On Veh. Tech.* Preprint available on arXiv.

[8] J. H. Manton, "Optimization algorithms exploiting unitary constraints", *IEEE Trans. Signal Process.* 50 (2002), no. 3, 635–650.

[9] J. Mangus, "On differentiating eigenvalues and eigenvectors", *Econometric Theory*, Vol. 1, No. 2 (Aug., 1985), pp. 179-191.

[10] Magnus, J. R. and H. Neudecker "Matrix differential calculus with applications in statistics and econometrics", *John Wiley & Sons*: Chichester/New York, reprinted 1990, pp. 198-200

[11] C. B. Papadias, "Globally convergent blind source separation based on a multiuser kurtosis maximization Criterion" *IEEE Trans. Signal Processing*, vol. 48, No. 12, pp. 3508-3519, Dec. 2000

[12] E. Polak, "Optimization: Algorithms and consistent approximations". New York: Springer-Verlag, 1997

[13] R. T. Krishnamachari, M. K. Varanasi "Interference alignemnt under limited feedback for MIMO ICs", *Proc. of ISIT 2009, Seoul, Korea*

[14] J. Thurkal, H. Bolcskei, "Interference Alingment under limited feedback", *Proc. ISIT 2009*, Seoul, Korea, pp. 1759-1763

[15] C. B. Papadias, "Globally convergent algorithms for bling source separation", *X Europ. Sign. Proc. Conf. (EUSIPCO 2000)*, Tampere, Finland, Sept 4-8 2000